\documentclass[12pt]{article}

\usepackage{amsmath,amssymb,amsthm,graphicx,latexsym}
\newcommand{\ed}{\end{document}}
\newcommand{\beq}{\begin{equation}}
\newcommand{\eeq}{\end{equation}}
\newcommand{\beqa}{\begin{eqnarray}}
\newcommand{\eeqa}{\end{eqnarray}}
\newcommand{\bc}{\begin{center}}
\newcommand{\ec}{\end{center}}
\newcommand{\vs}{\vspace}
\newcommand{\ba}{\begin{array}}
\newcommand{\ea}{\end{array}}

\def\({\left(}
\def\){\right)}

\begin{document}
\bc
\textbf{\large{Implication of Geodesic Equation in Generalized Uncertainty Principle framework}}\\
\vs{0.5cm}
Souvik Pramanik \footnote{souvick.in@gmail.com}\\
Physics and Applied Mathematics Unit,\\
Indian Statistical Institute,\\
203 B. T. Road, Kolkata 700108, India
\ec
\begin{abstract}
\emph{The generalized uncertainty principle (GUP) corrected modified relativistic particle model has been derived in curved space-time. From this modified model, the equation of motion (EM) has been constructed relativistically in terms of the affine parameter ($\lambda$) or proper time ($\tau$) and nonrelativistically in terms of coordinate time ($t$). In this context, the constraint analysis technique has been applied to get the EM. Interestingly, the EM obtained in both cases is the usual one. This result clearly indicates an important fact, that is, consistency of the equivalence principle in the GUP framework, and furthermore it can be concluded that with the GUP-corrected modified algebra it is impossible to get the GUP effect in point particle motion.}
\end{abstract}
\section{Introduction:}
Various approaches of quantum gravity such as string theory, doubly special relativity (DSR), black hole physics predict that there should exist a minimum measurable length at the order of the Planck length. By considering black hole gedanken experiment, it has been shown that gravity generates an uncertainty in determining the position of a black hole \cite{maggiore1,fabio,adler,luis}. Depending on such arguments, the well known Heisenberg's uncertainty principle has been modified to generalized uncertainty principle (GUP) \cite{maggiore2,kempf1,das1,das2}. This GUP incorporates the existence of a minimum measurable length and maximum observable momentum. Quantum mechanics has already been modified accordingly \cite{kempf2,kempf3}. All the quantum mechanical symmetries have been checked in this modified version. Recently, it has been shown that the GUP discloses a self-complete characteristic of gravity, namely, the possibility of masking any curvature singularity behind an event horizon \cite{nicolini}. In order to construct the quantum theory of gravity it is now required to check whether the GUP-corrected classical and the relativistic theory satisfy all the fundamental laws and principles of general relativity. By considering nonrelativistic GUP model, it has been shown that the equivalence principle (EP) is violated in the GUP framework \cite{tkachuk2}. But to verify the EP relativistically, here we first construct a GUP-corrected noncanonical modified structure of a point particle in curved space-time, and thereafter the equation of motion has been derived. Henceforth, considering a special nonrelativistic example by studying particle dynamics in terms of coordinate time $t$, we analyze the consistency of the \emph{equivalence principle} in the GUP framework.

Out of the few forms of the GUP \cite{kempf1,das1,das2}, we are interested in the model that has been proposed in Ref. \cite{kempf1}, because it is in a more general form than Ref. \cite{das1} and becomes \cite{das1} by linearizing with respect to the GUP parameter $\beta$. In order to get the classical structure corresponding to the quantum model \cite{kempf1}, we can take the help of the well known relation $\{X,P\}=\frac{[X,P]}{i\hbar}$, where $\{,\}$ stands for symplectic structure or Poisson Bracket and $[,]$ is the quantum commutator. Also comparing with Ref. \cite{tkachuk1,martin}, we can generate the relativistic four-dimensional form of our symplectic structures. Throughout this paper, we consider $(X,P)$ as canonical variables and $(x,p)$ as noncanonical variables \cite{kempf1}.

In order to build up modified structures, we resort to the following approach:

\emph{A GUP-corrected modified structure is constructed in terms of the noncanonical representation $(x,p)$, where the canonical representation $(X,P)$ satisfies all the usual known relations.}

This approach is quite precise to construct modified dynamics form GUP-corrected commutation relations \cite{kempf1,das1,das2}. This is because, in the GUP formalism, all these commutators have been written in terms of noncanonical variables $(x,p)$ where the canonical variables $(X,P)$ are known to satisfy the usual commutation relations $[X^\mu,P^\nu]=i\hbar \eta^{\mu\nu}$. Therefore, following the above approach we build up GUP-corrected noncanonical Lagrangian and Hamiltonian of a particle in Sec. 2. Afterward, in Sec. 3, we derive the equation of motion of that particle in an arbitrary reference frame (which may be curvilinear or accelerating) by applying a constraint analysis technique. From this equation of motion, we conclude about the consistency of equivalence principle in Sec. 4.

\section{GUP-corrected modified noncanonical structure}
\subsection{The noncanonical Lagrangian}
~~~In this section, first we build up the Lagrangian in flat space-time. Thereafter, from this flat space-time Lagrangian we construct a Lagrangian in curved space-time. The symplectic structures or the Poisson brackets corresponding to the GUP model \cite{kempf1} can be written in relativistic four-vector form in flat space-time as
\beq \{x^\mu,p^\nu\}=\frac{\beta p^2}{\sqrt{1+2\beta p^2}-1}\eta^{\mu\nu}+\beta p^\mu p^\nu, \label{xp-poi-bra}\eeq
\beq \{x^\mu,x^\nu\}=0~~~,~~~~\{p^\mu,p^\nu\}=0, \label{xxpp-poi-bra}\eeq
where the variables $(x,p)$ are noncanonical. We have considered the metric to be as usual $\eta^{\mu\nu}\equiv(-1,1,1,1)$. Now it has been shown that \cite{eune} the GUP-corrected Poisson brackets are the Dirac brackets of an extended system in which $x^{\mu}$ and $p^{\nu}$ are initially treated as independent configuration degrees of freedom with momenta $\Pi^{(x)\mu}$ and $\Pi^{(p)\nu}$, respectively. Second-class constraints are then imposed which eliminate $\Pi^{(x)\mu}$ and $\Pi^{(p)\nu}$ and turn $p^{\nu}$ into the momentum of $x^{\nu}$. Therefore, we can consider the above symplectic structures (\ref{xp-poi-bra}) and (\ref{xxpp-poi-bra}) as our Dirac brackets. With these Dirac brackets in hand, we derive the GUP-corrected point particle Lagrangian in flat space-time following the procedure presented in Ref. \cite{pramanik}. The procedure is going through the reverse direction of the conventional analysis
$$Lagrangian~\rightarrow~Constraints~\rightarrow~Dirac~brackets$$
or equivalently
$$ Symplectic~structure~\rightarrow~Symplectic~matrix~\rightarrow~Symplectic~brackets.$$
It is important to note that Dirac brackets and symplectic brackets are same. So, in this case, our path of analysis will be
$$Symplectic~brackets~\rightarrow~Symplectic~matrix~\rightarrow~Lagrangian.$$

In order to follow this path, it will be beneficial if we write down the mathematical method.

The generic structure of the symplectic brackets (SB) are of the form
\beq \{f,g\}_{SB}={\Gamma}^{\mu\nu}_{a b}(\partial_{a,\mu}f)(\partial_{b,\nu}g)\equiv\{f,g\}_{DB}
=\{f,g\}-\{f,\Phi^\mu_a\}\Sigma^{\mu\nu}_{a b}\{\Phi^\nu_b,g\},
\eeq
where $\partial_{a,\mu}=\frac{\partial}{\partial \eta^\mu_a}$, $\eta^\mu_1=x^\mu,~\eta^\mu_2=p^\mu$, and $\Phi^\mu_a$ are second-class constraints. Now the inverse of this $\Sigma$ matrix provides the constraint matrix $\Sigma^{a b}_{\mu\nu}=\{\Phi^a_\mu,\Phi^b_\nu\}$. If $\Pi^{(x)}_\mu=\frac{\partial L}{\partial \dot{x}^\mu}$ and $\Pi^{(p)}_\nu=\frac{\partial L}{\partial \dot{p}^\nu}$ are the momenta corresponding to the variable $x$ and $p$ respectively, that satisfy the Poisson brackets $\{x_\mu,\Pi^{(x)}_\nu\}=\eta_{\mu\nu}$ and $\{p_\mu,\Pi^{(p)}_\nu\}=\eta_{\mu\nu}$, then form the constraint matrix $\{\Phi^a_\mu,\Phi^b_\nu\}$, we can make a judicious choice of the constraints containing the momenta $\Pi^{(x)}_\mu$ and $\Pi^{(p)}_\nu$. The presence of these momentum terms in the constraints is required to construct a Lagrangian from constraint structures.

Now following this method first the symplectic matrix that can be formed from the above symplectic structure (\ref{xp-poi-bra}) and (\ref{xxpp-poi-bra}) is
\begin{equation}
\Sigma^{\mu\nu}=\left[ {\begin{array}{cc}
0 & \Lambda \eta^{\mu\nu}+\beta p^\mu p^\nu \\
-\Lambda \eta^{\mu\nu}-\beta p^\mu p^\nu & 0 \\
\end{array} }\right],
\label{sim-mat}
\end{equation}
where $\Lambda=\frac{\beta p^2}{\sqrt{1+2\beta p^2}-1}$. The inverse of this symplectic matrix yields the matrix of constraint brackets
\begin{equation}
\Sigma_{\mu\nu}=\left[ {\begin{array}{cc}
0 & -\frac{\eta_{\mu\nu}}{\Lambda}+\frac{\beta p_\mu p_\nu}{\Lambda^2\sqrt{1+2\beta p^2}} \\
\frac{\eta_{\mu\nu}}{\Lambda}-\frac{\beta p_\mu p_\nu}{\Lambda^2\sqrt{1+2\beta p^2}} & 0 \\
\end{array} }\right]\equiv\{\Phi^i_\mu,\Phi^j_\nu\}.
\label{sim-inv-mat}
\end{equation}
Then from (\ref{sim-inv-mat}) the structure of constraints that can be formed is
\beq \Phi^1_\mu=\Pi^{(x)}_\mu-\frac{p_\mu}{\Lambda}\approx0,~~~~\Phi^2_\nu=\Pi^{(p)}_\nu\approx 0, \label{cons1} \eeq
where $\Pi^{(x)}_\mu=\frac{\partial L}{\partial \dot{x}^\mu}$ and $\Pi^{(p)}_\nu=\frac{\partial L}{\partial \dot{p}^\nu}$ satisfy the Poisson brackets $\{x_\mu,\Pi^{(x)}_\nu\}=\eta_{\mu\nu}$, and $\{p_\mu,\Pi^{(p)}_\nu\}=\eta_{\mu\nu}$. With this constraint structure (\ref{cons1}) in hand, one can derive the Dirac bracket between the variables $(x,p)$ and check the consistency of the constraints (\ref{cons1}). Now (\ref{cons1}) implies $\Pi^{(x)}_\mu=\frac{\partial L}{\partial \dot{x}^\mu}=\frac{p_\mu}{\Lambda}$ and $\Pi^{(p)}_\nu=\frac{\partial L}{\partial \dot{p}^\nu}=0$. Integrating these two relations we have the Lagrangian that is compatible with (\ref{cons1}):
\beq L_{NC}(x(\lambda))=\frac{(\eta_{\mu\nu}p^\mu\frac{dx^\nu}{d\lambda})}{\Lambda}+\emph{v}(\lambda)(f(p^2)+m^2c^2), \label{lag1}\eeq
where $\emph{v}(\lambda)$ is the Lagrange multiplier. Here $\lambda$ is the affine parameter, which is linearly related to proper time $\tau$ by $\lambda=a+b\tau$ \cite{carroll}, for any arbitrary constant $a$ and $b$. One can construct the function $f(p^2)$ as $f(p^2)=(\frac{p^\mu}{\Lambda})^2$ \cite{pramanik}.

Another structure of constraint that can be formed from (\ref{sim-inv-mat}) is $\Phi_\mu^1=\Pi_\mu^{(x)}\approx0$ and $\Phi_\nu^2=\Pi_\nu^{(p)}+\frac{x_\nu}{\Lambda}-\frac{\beta(xp) p_\nu}{\Lambda^2\sqrt{1+2\beta p^2}}\approx0$ \cite{pramanik}. The corresponding Lagrangian that can be constructed is $L_{NC}(x(\lambda))=-\frac{(x\dot{p})}{\Lambda}+\frac{\beta(xp)(p\dot{p})}{\Lambda^2\sqrt{1+2\beta p^2}}+\emph{v}(\lambda)(f(p^2)+m^2c^2)$ \cite{pramanik}. Though this Lagrangian and (\ref{lag1}) seem to be completely different, actually one can be achieved from another just by doing one time integration by parts on the action integral. In other words, these two are equivalent. This can be shown as follows: the action corresponding to the Lagrangian (\ref{lag1}) is [by dropping the last term of (\ref{lag2})]
\beq S=\int L_{NC} d\lambda=\int \frac{\eta_{\mu\nu}p^\mu \frac{d x^\nu}{d\lambda}}{\Lambda} d\lambda. \label{action1}
\eeq
Then by performing integration by parts on the right-hand side of (\ref{action1}) we get the Lagrangian \cite{pramanik}
\beq S=\int\left\{-\frac{(\eta_{\mu\nu}x^\mu\dot{p}^\nu)}{\Lambda}+\frac{\beta(\eta_{\mu\nu}x^\mu p^\nu)(\eta_{\mu\nu}p^\mu \dot{p}^\nu)}{\Lambda^2\sqrt{1+2\beta p^2}}\right\}d\lambda.
\label{action2}\eeq

We are interested in the form (\ref{lag1}), because the Lagrangian (\ref{lag1}) is much more suitable than (\ref{action2}) for constructing a Lagrangian in curved space-time. Before going into detail, let us consider first for simplicity the lowest nontrivial lowest order of $\beta$, i.e. the first order of $O(\beta)$. Up to this first order of $O(\beta)$, the Lagrangian (\ref{lag1}) gets its $O(\beta)$ structure as
\beq L_{NC}(x(\lambda))\approx\left(1-\frac{\beta}{2}p^2\right)\eta_{\mu\nu}p^\mu \frac{d x^\nu}{d\lambda}+\emph{v}(\lambda)(p^2-\beta (p^2)^2+ m^2c^2). \label{lag2} \eeq
Let us define $A=\left(1-\frac{\beta}{2}p^2\right)$. One can check the consistency of the Lagrangian (\ref{lag2}) by deriving the Poisson bracket [$\{B,C\}_{PB}=\sum(\frac{\partial B}{\partial x^\mu}\frac{\partial C}{\partial \Pi^{(x)}_\mu}-\frac{\partial C}{\partial x^\mu}\frac{\partial B}{\partial \Pi^{(x)}_\mu})$, where $\Pi^{(x)}_\mu=\frac{\partial L(O(\beta))}{\partial \dot{x}^\mu}$] between $x^\mu$ and $p^\nu$. Once again, it can be shown that this Lagrangian (\ref{lag2}) is equivalent to the Lagrangian that can be derived from (\ref{action2}) just by considering the first order of $\beta$:
\beq L_{NC}(x(\lambda))=-\left(1-\frac{\beta}{2}p^2\right)(\eta_{\mu\nu} x^\mu \dot{p}^\nu)+\beta (\eta_{\mu\nu} x^\mu p^\nu)(\eta_{\mu\nu} p^\mu \dot{p}^\nu) \label{lag4} \eeq

This form of Lagrangian (\ref{lag4}) is too complicated to get a Lagrangian in curved space-time. This is due to the presence of $\dot{p}$ terms in (\ref{lag4}). The $\dot{p}$ terms can arise in the Lagrangian only by performing integration by parts on the usual Lagrangian containing $\dot{x}$ terms. In curved space-time, whenever $\eta_{\mu\nu}$ is replaced by $g_{\mu\nu}$, then this integration by parts generates extra terms like the derivative of $g_{\mu\nu}$ which are difficult to guess from a flat space-time Lagrangian like (\ref{lag4}). But since the Lagrangian (\ref{lag2}) contains as usual only the $\dot{x}$ term and not $\dot{p}$, then a Lagrangian in curved space-time can easily be obtained just by replacing all $\eta_{\mu\nu}$ by $g_{\mu\nu}$:
\beq
L_{NC}(x(\lambda))=\left(1-\frac{\beta}{2}p^2\right)g_{\mu\nu}p^\mu \frac{d x^\nu}{d\lambda}. \label{lag3}
\eeq
This is our point particle Lagrangian in curved space-time. This Lagrangian is as usual of the form $L=g_{\mu\nu}\dot{x}^\mu \Pi^{(x)\nu}$. But one cannot start just by replacing all canonical variables by noncanonical ones in canonical Lagrangian, because we have only GUP-corrected Poisson brackets in hand and we have to build up a Lagrangian compatible with it.

The equivalent form of this Lagrangian (\ref{lag3}) that contains $\dot{p}$ terms can be obtained by performing integration by parts on the corresponding action which yields
\beq
L_{NC}(x(\lambda))=-A(g_{\mu\nu}x^\mu\dot{p}^\nu-\partial_\gamma(g_{\mu\nu})\dot{x}^\gamma p^\mu x^\nu)+\beta (g_{\mu\nu}x^\mu p^\nu)(g_{\mu\nu}p^\mu\dot{p}^\nu)+\frac{\beta}{2}(g_{\mu\nu}x^\mu p^\nu)\partial_\gamma(g_{\mu\nu})\dot{x}^\gamma p^\mu p^\nu. \label{lag5}
\eeq
This form of the Lagrangian is difficult to construct from the Lagrangian (\ref{lag4}). In Ref. \cite{ghosh}, the curved space-time Lagrangian (constructed from a flat space-time Lagrangian) does not contain all terms of the GUP-corrected point particle Lagrangian (\ref{lag5}) in curved space-time or equivalently (\ref{lag3}).
\subsection{The noncanonical Hamiltonian}
~~~To the first order of $O(\beta)$ the Poisson brackets (\ref{xp-poi-bra}) can be written in the curved space-time background as
\beq \{x^\mu,p^\nu\}=\left(1+\frac{\beta}{2}p^2\right)g^{\mu\nu}+\beta p^\mu p^\nu, \label{xp-poi-bra1}
\eeq
where $g^{\mu\nu}\equiv(-,+,+,+)$. The relation between the noncanonical variables $(x,p)$ and canonical variables $(X,P)$ can be constructed from the above bracket (\ref{xp-poi-bra1}):
\beq x^\mu=X^\mu~,~~p^\nu=P^\nu\left(1+\frac{\beta}{2} P^2\right). \label{modi-x-p} \eeq
Since the canonical momentum $P$ satisfies the dispersion relation $P^2+m^2 c^2=0$, we get our modified dispersion relation as \cite{bibhas}
\beq p^2-\beta (p^2)^2+m^2c^2=0. \label{dis-rel} \eeq
Now to derive the Hamiltonian we take help of the technique presented in Ref. \cite{hans}. First of all, differentiating the Lagrangian (\ref{lag3}) with respect to $\frac{d x^\mu}{d\lambda}$ gives
$\Pi^{(x)}_\mu(\lambda)=\frac{\partial L}{\partial (\frac{d x^\mu}{d\lambda})}=g_{\mu\gamma}p^\gamma\left(1-\frac{\beta}{2}p^2\right)$,
which provides
\beq H_1=g_{\mu\gamma}\frac{d x^\mu}{d\lambda} \Pi^{(x)\gamma}(\lambda)-L=g_{\mu\gamma}\frac{d x^\mu}{d\lambda} p^\gamma\left(1-\frac{\beta}{2}p^2\right)-L=0. \label{ham1} \eeq
As we have first-class primary constraint (\ref{dis-rel}), then the total Hamiltonian can be written as
\beq H=H_1+\emph{v}(\lambda)(p^2-\beta (p^2)^2+m^2c^2), \label{ham2} \eeq
where $\emph{v}(\lambda)$ is an unknown function that has to be determined. $H$ correctly generates Hamilton's equation of motion with respect to the parameter $\lambda$:
\beq \dot{x}^\mu \equiv \{x^\mu,H\}=\frac{\partial H}{\partial\Pi^{(x)}_\mu}
=2\emph{v}(\lambda)p^\mu\left(1-\frac{\beta}{2}p^2\right). \label{x-mu-dot} \eeq
The last relation of (\ref{modi-x-p}) implies $P^\mu=p^\mu(1-\frac{\beta}{2}p^2)$. Since the canonical Lagrangian is $L_C=-mc\sqrt{-g_{\mu\nu}\frac{d X^\mu}{d\lambda}\frac{d X^\nu}{d\lambda}}$, where the canonical momentum $P$ is related to $\dot{X}$ by the relation $P^\mu=\frac{m c \frac{d X^\mu}{d\lambda}}{\sqrt{-g_{\rho\sigma}\frac{d X^\rho}{d\lambda}\frac{d X^\sigma}{d\lambda}}}$, then one can write $p^\mu(1-\frac{\beta}{2}p^2)=\frac{m c \frac{dx^\mu}{d\lambda}}{\sqrt{-g_{\rho\sigma}\frac{d x^\rho}{d\lambda}\frac{d x^\sigma}{d\lambda}}}$. Substituting this into (\ref{x-mu-dot}) gives the coefficient $\emph{v}(\lambda)$  as $\emph{v}(\lambda)=\frac{1}{2 m c}\sqrt{-g_{\rho\sigma}\frac{d x^\rho}{d\lambda}\frac{d x^\sigma}{d\lambda}}$. Therefore, our relativistic Hamiltonian becomes
\beq H=\frac{1}{2mc}\left(p^2-\beta (p^2)^2+m^2c^2\right)\sqrt{-g_{\rho\sigma}\frac{d x^\rho}{d\lambda}\frac{d x^\sigma}{d\lambda}}. \label{ham3} \eeq
This is the final form of our GUP-corrected relativistic Hamiltonian. However, the $\dot{x}$ term present in (\ref{ham3}) can be scaled by choosing a suitable gauge constraint, but in order to study the dynamics from an arbitrary reference frame we cannot do this.

\section{Derivation of Dirac brackets and equation of motion}
To derive the equation of motion by the constraint analysis technique, first of all it is essential to find the Dirac brackets between the noncanonical variables $x$ and $p$ for such a constraint system. In this section, first we concentrate on this.

Now the momentums corresponding to the variables $x$ and $p$ obtained from the above Lagrangian (\ref{lag3}) are
\beq \Pi^{(x)}_\mu(\lambda)=\frac{\partial L}{\partial (\frac{d x^\mu}{d\lambda})}=\left(1-\frac{\beta}{2}p^2\right)
g_{\mu\lambda}p^\lambda~,~~~~\Pi_\mu^{(p)}(\lambda)=\frac{\partial L}{\partial (\frac{d p^\mu}{d\lambda})}=0. \label{mom1}\eeq
The structure of constraints can be constructed from the above momenta:
\beq \Phi^1_\mu=\Pi^{(x)}_\mu(\lambda)-\left(1-\frac{\beta}{2}p^2\right)g_{\mu\gamma}p^\gamma\approx 0~,
~~~~\Phi^1_\nu=\Pi^{(p)}_\nu(\lambda)\approx 0, \label{cons2} \eeq
which yields the constraint matrix
\begin{eqnarray}
\{\Phi^i_\mu,\Phi^j_\nu\} &=& \left[ {\begin{array}{cc}
A Q_{\mu\nu}-\frac{\beta}{2}M_{\mu\nu} & -A g_{\mu\nu}+\beta p_\mu p_\nu \\
A g_{\mu\nu}-\beta p_\mu p_\nu & 0 \\
\end{array} }\right] \nonumber\\
&=& \left[ {\begin{array}{cc}
A Q_{\mu\nu} & -A g_{\mu\nu}+\beta p_\mu p_\nu \\
A g_{\mu\nu}-\beta p_\mu p_\nu & 0 \\
\end{array} }\right]
-\frac{\beta}{2}\left[ {\begin{array}{cc}
M_{\mu\nu} & 0 \\
0 & 0 \\
\end{array} }\right]
=[\mathcal{A}]-\frac{\beta}{2}[\mathcal{B}]
\label{con-mat-1}
\end{eqnarray}
where $Q_{\mu\nu}=(\partial_\mu(g_{\nu c})-\partial_\nu(g_{\mu c}))p^c$ and
$M_{\mu\nu}=(p_\nu \partial_\mu(g_{\rho\sigma})-p_\mu \partial_\nu(g_{\rho\sigma}))p^\rho p^\sigma$. The inverse of this
constraint matrix to the first order of $O(\beta)$ can be written as
$\{\Phi^i_\gamma,\Phi^j_\mu\}^{-1}=\left([\mathcal{A}]^{-1}\right)^{\gamma\mu}
+\frac{\beta}{2}\left([\mathcal{A}]^{-1}\right)^{\gamma \rho}\left([\mathcal{B}]\right)_{\rho\sigma}
\left([\mathcal{A}]^{-1}\right)^{\sigma \mu},$
which after using the definition of Dirac brackets provides the Dirac brackets between the noncanonical variables $(x,p)$ as
\begin{eqnarray}
\left[x^\gamma,p^\mu\right]_D &=& \frac{1}{A}g^{\gamma\mu}+\frac{\beta}{A(A-\beta p^2)}p^\gamma p^\mu
\approx \left(1+\frac{\beta}{2}p^2\right)g^{\gamma\mu}+\beta p^\gamma p^\mu,\nonumber\\
\left[p^\gamma,p^\mu\right]_D &=& \frac{1}{A}g^{\gamma \rho}g^{\mu \sigma}Q_{\rho\sigma}
-\frac{\beta}{A(A-\beta p^2)}(p^\gamma g^{\mu \rho}-p^\mu g^{\gamma \rho})Q_{\rho\sigma}p^\sigma
-\frac{\beta}{2 A^2}g^{\gamma \rho}g^{\mu\sigma}M_{\rho\sigma}\nonumber\\
&\approx& \left(1+\frac{\beta}{2}p^2\right)g^{\gamma \rho}g^{\mu \sigma}Q_{\rho\sigma}-\beta(p^\gamma g^{\mu \rho}-p^\mu g^{\gamma\rho})Q_{\rho\sigma}p^\sigma-\frac{\beta}{2}g^{\gamma \rho}g^{\mu \sigma}M_{\rho\sigma},\nonumber\\
\left[x^\gamma,x^\mu\right]_D &=& 0. \label{dir-bra}
\end{eqnarray}
Interestingly, the Dirac bracket between $x$ and $p$ is the same as their Poisson bracket (\ref{xp-poi-bra1}). By using these Dirac brackets (\ref{dir-bra}) and the above Hamiltonian (\ref{ham3}), the equations of motion are obtained from
\beq \frac{d x^\mu}{d\lambda}=[x^\mu,H]_D~,~~~~\frac{d p^\mu}{d\lambda}=[p^\mu,H]_D. \label{ham-equ} \eeq
From the first equation of (\ref{ham-equ}), we have
\beq \frac{d x^\mu}{d\lambda}=p^\mu\left(1-\frac{\beta}{2}p^2\right)\frac{\sqrt{-g_{\rho\sigma}\frac{d x^\rho}{d\lambda}\frac{d x^\sigma}{d\lambda}}}{m c}. \label{dotx} \eeq
It is easy to verify the dispersion relation (\ref{dis-rel}) from (\ref{dotx}). Again to the first order of $\beta$, Eq. (\ref{dotx}) can be written as
\beq p^\mu=\frac{mc\frac{dx^\mu}{d\lambda}}{\sqrt{-g_{\rho\sigma}\frac{d x^\rho}{d\lambda}\frac{d x^\sigma}{d\lambda}}}
\left(1+\frac{\beta}{2}p^2\right). \label{p-mu} \eeq
Now differentiating (\ref{p-mu}) with respect to the affine parameter $\lambda$, we get
\beq \frac{d p^\mu}{d\lambda}=\frac{m c\frac{d^2 x^\mu}{d\lambda^2}}{\sqrt{-g_{\rho\sigma}\frac{d x^\rho}{d\lambda}\frac{d x^\sigma}{d\lambda}}}\times\left(1+\frac{\beta}{2}p^2\right)+\frac{\beta m c \frac{dx^\mu}{d\lambda}}{2\sqrt{-g_{\rho\sigma}\frac{d x^\rho}{d\lambda}\frac{d x^\sigma}{d\lambda}}}\frac{d(p^2)}{d\lambda}. \label{ddotx} \eeq
To obtain $\frac{d(p^2)}{d\lambda}$ we can take the help of the relation $\frac{d(p^2)}{d\lambda}=[p^2,H]_D$, which yields $\frac{d(p^2)}{d\lambda}=0$. Now from the second equation of (\ref{ham-equ}), we have
\beq \frac{d p^\mu}{d\lambda}=\frac{(1-2\beta p^2)}{mc}\left[-\frac{1}{2}\partial_\gamma(g_{\nu\delta})p^\nu p^\delta
[x^\gamma,p^\mu]_D+p_\nu[p^\mu,p^\nu]_D\right]\sqrt{-g_{\rho\sigma}\frac{d x^\rho}{d\lambda}\frac{d x^\sigma}{d\lambda}}.
\label{dotp} \eeq
Using the Dirac brackets (\ref{dir-bra}) and replacing all $p$ by (\ref{p-mu}), the right-hand side of (\ref{dotp}) to the first order of $O(\beta)$ becomes $\frac{d p^\mu}{d\lambda}=-\frac{mc\left(1-\frac{\beta}{2}m^2c^2\right)}{\sqrt{-g_{\rho\sigma}\frac{d x^\rho}{d\lambda}\frac{d x^\sigma}{d\lambda}}}\Gamma^\mu_{\nu\delta}\frac{d x^\nu}{d\lambda}\frac{d x^\delta}{d\lambda}$. Comparing this with (\ref{ddotx}) immediately gives the same geodesic equation as usual:
\beq \frac{d^2 x^\mu}{d\lambda^2}+\Gamma^\mu_{\nu\gamma}\frac{d x^\nu}{d\lambda}\frac{d x^\gamma}{d\lambda}=0. \label{geo-equ} \eeq
Thus in the noncanonical representation $(x,p)$ the geodesic equation remains unchanged. In other words, the particle dynamics in curved space-time do not change by considering the usual algebra (\ref{modi-x-p}) obtained from the GUP.
\section{Analytic discussion on the above Results}
In this section, we discuss the consistency of the equivalence principle in the GUP framework. Let us consider a specific example by changing our independent parameter (affine parameter $\lambda$) to coordinate time $t$ and take the nonrelativistic limit from (\ref{lag3}).

But before that it is required to note that the canonical Lagrangian $L_C$ and the noncanonical one $L_{NC}$ are different whenever they are written in terms of $(\dot{X},P)$ and $(\dot{x},p)$, respectively. But their form becomes the same if we write them in terms of only $\dot{X}$ and $\dot{x}$, respectively. This can be achieved by replacing all $p$ in (\ref{lag3}) by (\ref{p-mu}), and as a result $L_{NC}$ becomes
\beq L_{NC}(x(\lambda))=-mc\sqrt{-g_{\mu\nu}\frac{d x^\mu}{d\lambda}\frac{d x^\nu}{d\lambda}}. \label{L-dotx}\eeq
The reason behind the noncanonical Lagrangian (\ref{L-dotx}) having the same form as canonical Lagrangian  $L_{C}=-mc\sqrt{-g_{\mu\nu}\frac{d X^\mu}{d\lambda}\frac{d X^\nu}{d\lambda}}$ is that we have taken $x=X$ through (\ref{modi-x-p}). On the other hand, since we have modified the momentum part $P$ (\ref{modi-x-p}), in order to get the GUP effect we have to go to the Hamiltonian formalism and derive Hamilton's equations of motion.

First, we write down the Lagrangian (\ref{L-dotx}) in terms of $\frac{dx(t)}{dt}$ in order to study the dynamics in coordinate time $t$. This yields $L_{NC}(x(t))$ as
\beq L_{NC}(x(t))=-m c^2\sqrt{-g_{00}}\sqrt{1+\frac{g_{ii}(\frac{d x^i}{d t})^2}{g_{00}c^2}},\label{L-dotx-t} \eeq
where $g_{\mu\nu}$ is considered as a diagonal matrix to avoid the complexity of calculation. The corresponding Hamiltonian is [see the Appendix]
\beq H_{NC}(p)=m c^2 \sqrt{-g_{00}}\sqrt{1+\frac{g_{ii}(p^i)^2}{m^2c^2(1-\beta m^2 c^2)}}. \label{ham4} \eeq

Now we want to study the dynamics of a freely falling particle in Earth's uniform gravitational field from a coordinate system fixed on Earth. In this context, we consider $g_{ii}=g_{rr}$ only. In nonrelativistic limit, the above Hamiltonian (\ref{ham4}) becomes
\beq H=m c^2+\frac{p_r^2}{2m(1-\beta m^2 c^2)}+m\Phi(r), \label{ham5} \eeq
where we have used the approximation $g_{\mu\mu}=\eta_{\mu\mu}+h_{\mu\mu}$, $[h_{\mu\mu}\ll 1,~\mu=(t,~r)]$ with $h_{00}=h_{rr}=-\frac{2\Phi}{c^2}$ and neglect the terms containing $O(\frac{1}{c^2})$. Here $\Phi$ is the gravitational potential. With such kinds of approximation the Poisson bracket (\ref{xp-poi-bra1}) can be written as $\{r,p_r\}=(1-\frac{\beta}{2}m^2 c^2)+\beta m^2 \dot{r}^2$. Since $\beta=\frac{\beta_0}{M_{pl}^2 c^2}$ \cite{das1}, the last term of the Poisson bracket is $O(\frac{1}{c^2})$, which can be further neglected. Thus the Poisson bracket finally becomes
\beq \{r,p_r\}=\left(1-\frac{\beta}{2}m^2 c^2\right). \label{poi-bra} \eeq
Hamilton's equations of motion $\frac{dr}{dt}=\{r,H\}$ and $\frac{dp_r}{dt}=\{p_r,H\}$ then give the differential equation for the motion as
\beq \ddot{r}=-\nabla\Phi. \label{equ-mot}\eeq
This is just the usual Newton's law of gravity. Therefore, in terms of coordinate time $t$ we have again reached the usual equation of motion. It is clear from (\ref{geo-equ}) and (\ref{equ-mot}) that the equivalence principle is consistent in the GUP framework. We have reached to (\ref{equ-mot}) depending on nonrelativistic Hamiltonian (\ref{ham5}) and Poisson bracket (\ref{poi-bra}). Interestingly, this Hamiltonian (\ref{ham5}) incorporates GUP corrections. But the GUP corrections are not present in the Hamiltonian \cite{tkachuk2}. By reason of nonrelativistic limit [neglecting $O(\frac{1}{c^2})$ terms] the Poisson bracket (\ref{poi-bra}) does not contain any $\dot{x}$ terms, whereas the $\dot{x}$ terms are present in the Poisson bracket \cite{tkachuk2}. The presence of these $\dot{x}$ terms in the Poisson bracket therein has brought off the dependence of test particle mass $m$ in the motion \cite{tkachuk2}. But the result we have obtained here is that the equation of motion remains unchanged whenever one studies the dynamics in terms of the affine parameter $(\lambda)$ or proper time $(\tau)$ or coordinate time $t$.
\section{Conclusion:}
Going through a consistent way of constraint analysis, we have derived here a GUP-corrected modified point particle Lagrangian (\ref{lag3}), Hamiltonian (\ref{ham3}), and thereafter the equation of motion (\ref{geo-equ}) in curved space-time. These structures are derived here by taking the affine parameter $(\lambda)$ (or proper time $\tau$) as an independent variable. Henceforth, the relativistic and nonrelativistic GUP-corrected point particle Hamiltoniana (\ref{ham4}) and (\ref{ham5}) are also obtained in terms of the coordinate time $t$ (of an arbitrary reference frame). Such a point particle dynamics has also been studied in the context of $\kappa$-Minkowski space-time in Ref. \cite{harikumar}. In Ref. \cite{tkachuk2} it has been shown that the equivalence principle in violated in the GUP framework whenever particle dynamics is studied nonrelativistically by taking $t$ as an independent parameter. But the important result we have obtained here is that, whenever one studies the particle dynamics from any arbitrary reference frame (whatever this may be) by taking affine parameter $\lambda$ or proper time $\tau$ or coordinate time $t$ as a dynamical parameter, the same geodesic equation will reproduced. This result clearly indicates consistency of the equivalence principle in the GUP framework. Consistency of the equivalence principle gives some support to construct quantum theory of gravity with the GUP. But with the proposed algebra (\ref{modi-x-p}) in hand, it is impossible to get the GUP effect classically in point particle dynamics. More specifically, only by modifying the momentum part $P$ one cannot obtain any GUP effect in single particle motion. This analysis brings up the thought of whether it is possible to get GUP effects by deriving other algebras from GUP models \cite{kempf1,das1}. For instance, one possibility is that one can study GUP effects by modifying the curvature tensor part and building up modified algebra. For this, we have to look at the near future.\\
{\bf Acknowledgement:} I thank to Professor Subir Ghosh for his useful discussion.
\section{Appendix}
The momenta corresponding to $L_{NC}(x(t))$ are $\Pi^{(x)}_j(t)=\frac{\partial L_{NC}(x(t))}{\partial (\frac{d x^j}{d t})}=\frac{m g_{jj}\frac{d x^j}{d t}}{\sqrt{-g_{00}}\sqrt{1+\frac{g_{ii}(\frac{d x^i}{d t})^2}{g_{00}c^2}}}$, using which we get the corresponding Hamiltonian as
\beq H_{NC}(x(t))=\frac{m c^2 \sqrt{-g_{00}}}{\sqrt{1+\frac{g_{ii}(\frac{d x^i}{d t})^2}{g_{00}c^2}}}.\label{ham12}\eeq
Now, since $\Pi^{(x)}_j(\lambda)=\frac{\partial L_{NC}(x(\lambda))}{\partial(\frac{d x^j}{d\lambda})}=\frac{mc g_{\mu j}\frac{d x^\mu}{d\lambda}}{\sqrt{-g_{\mu\nu}\frac{d x^\mu}{d\lambda}\frac{d x^\nu}{d\lambda}}}$ can be written as $\frac{m g_{ij}\frac{d x^i}{d t}}{\sqrt{-g_{00}}\sqrt{1+\frac{g_{ii}(\frac{d x^i}{d t})^2}{g_{00}c^2}}}$, then comparing $\Pi^{(x)}_i(\lambda)$ with (\ref{mom1}) we get
\beq \frac{m g_{ij}\frac{d x^i}{d t}}{\sqrt{-g_{00}}\sqrt{1+\frac{g_{ii}(\frac{d x^i}{d t})^2}{g_{00}c^2}}}=g_{ij}p^i\left(1-\beta\frac{p^2}{2}\right). \label{xdot-p-rel} \eeq
A little calculation leads to $\left(1+\frac{g_{ii}(\frac{d x^i}{d t})^2}{g_{00}c^2}\right)=\frac{1}{1+\frac{g_{ii}(p^i)^2}{m^2c^2(1-\beta m^2 c^2)}}$ , which finally yields modified Hamiltonian (\ref{ham4}).

\end{document}